\documentclass[%
    aip,
    jcp,
    amsmath,
    amssymb,
    twocolumn,
    superscriptaddress,
    10pt]{revtex4-1}
\usepackage{dcolumn}
\usepackage{graphicx} 
\usepackage{bm}
\usepackage{longtable}
\usepackage[english]{babel}
\usepackage{color}

\begin{document}

\title{Conformational dynamics modulating electron transfer}
\author{Dmitry V.\ Matyushov}
\affiliation{School of Molecular Sciences and Department of Physics, Arizona State University, PO Box 871504, Tempe, Arizona 85287}
\email{dmitrym@asu.edu}

\begin{abstract}
	Diffusional dynamics of the donor-acceptor distance in electron-transfer reactions are responsible for the appearance of a new time scale of diffusion over the distance of falloff of electronic tunneling. The distance dynamics compete with the medium polarization dynamics in the solvent-controlled electron transfer kinetics. A new solution incorporating the medium and donor-acceptor dynamics into the electron-transfer rate constant is proposed. The pre-exponential factor of the rate constant switches between a distance-independent solvent-controlled regime and exponential distance decay.  The crossover between two regimes is controlled by an effective relaxation time slowed down by a factor exponentially depending on the variance of the donor-acceptor displacement. Flexible donor-acceptor complexes must show a greater tendency for solvent-controlled electron transfer. Electron-transfer energy chains are best designed by placing the redox cofactors near the crossover distance.  
\end{abstract}

\maketitle

\section{Introduction}
Models of solvent (friction) control of electron transfer\cite{Zusman:80,Friedman:1982eu,Sumi:86,Hynes:86,Rips:1987qf,Yan:1988mz}  allow the pre-exponential factor of the reaction rate constant to be controlled by either the tunneling probability or by the relaxation time of the solvent models coupled to the reaction coordinate. This general formulation yields the rate constant of electron transfer $k_\text{ET}$ as the ratio of the nonadiabatic, golden rule rate constant $k_\text{NA}$ and the correction factor $1+g$
\begin{equation}
k_\text{ET}=  \frac{ k_\text{NA}}{1+g} . 
\label{1}
\end{equation}

The nonadiabatic rate constant $k_\text{NA}$ is given by the quantum mechanical perturbation (golden rule) expression averaged over statistical fluctuations of the reaction coordinate $X$
\begin{equation}
  k_\text{NA} = \frac{2\pi V_e^2}{\hbar}\left\langle\delta (X)\right\rangle. 
  \label{2}
\end{equation}
Here, the electron-transfer coupling $V_e$ refers to the equilibrium distance between the electron donor and acceptor (subscript ``e''). The crossover parameter 
\begin{equation}
g\propto \tau_x V_e^2  
\label{6}
\end{equation}
in the denominator of Eq.\ \eqref{1} is proportional to the product of $V_e^2$ and the relaxation time $\tau_x$ of the reaction coordinate $X$ supplied experimentally or computationally by the Stokes-shift dynamics.\cite{Zwan:85,Maroncelli:93} 

The reaction coordinate $X$ for electron transfer is specified by the donor-acceptor energy gap, which is the vertical (at frozen nuclei) energetic separation between the donor and acceptor energy levels.\cite{Lax:52,Warshel:1982wc} Its average value, $X_0=\lambda+\Delta G$, is a sum of the electron-transfer reorganization energy  $\lambda$  and the reaction free energy $\Delta G$.\cite{MarcusSutin} The tunneling state, in which electron transfer occurs, is reached at $X=0$; this condition is imposed by the delta function in Eq.\ \eqref{2}. If $X$ is a Gaussian fluctuating variable, the reaction activation barrier is fully determined by two statistical moments: the average $X_0$ and the variance 
\begin{equation}
  \sigma_X^2=\langle(\delta X)^2\rangle = 2\lambda k_\text{B}T,  
  \label{3}
\end{equation}
where $\delta X= X-X_0$ and $\lambda$ is the Marcus reorganization energy of electron transfer.\cite{MarcusSutin}    

At small values of $g$, the reaction is in the nonadiabatic tunneling regime. When $g>1$, the rate is switched to the dynamical control with $k_\text{ET}\propto \tau_x^{-1}$ corresponding to Kramers' activated kinetics.\cite{Kramers:1940wm,FrauenfelderWolynes:85} The medium dynamics can be complex and not reducible to a single relaxation time,\cite{Yan:1988mz} which makes the population dynamics non-exponential and complicates the formalism.\cite{Sumi:86,Gayathri:96} Nevertheless, both the simplified result in Eq.\ \eqref{1} and its more advanced analogs fundamentally describe the competition of the dissipative relaxation time and the tunneling time $\sim \hbar/V_e$ in the region where electronic terms representing the reactants and products for the electron-transfer reaction cross to form the reaction activation barrier.\cite{FrauenfelderWolynes:85} The effect of medium dynamics enters these formalisms through the dynamics of the coordinate $X$ (Stokes-shift dynamics\cite{Zwan:85,Maroncelli:93}).  

The medium dynamics can also modulate the tunneling probability. This non-Condon effect enters the description of electron transfer in terms of the electronic coupling $V(R)$, which can be affected either by fluctuating energy levels of the medium in superexchange electronic coupling\cite{Zhang:2014hn} or through fluctuations of the donor-acceptor distance $R$. If dynamics are not involved, one takes the statistical average in Eq.\ \eqref{2} over the distribution of donor-acceptor distances $R$ affecting the electronic coupling through an exponential decay\cite{Hopfield:1974uq,Winkler:2014ds} 
\begin{equation}
V(R) = V_e e^{-\tfrac{1}{2}\gamma \delta R}  ,
\label{4}
\end{equation}
where  $V_e$, as in Eq.\ \eqref{2}, is defined at the equilibrium distance $R_e$, $\gamma$ is the electron tunneling decay parameter, and $\delta R= R-R_e$ is the distance displacement. Applying Gaussian statistics of $R$ leads to the appearance of a temperature-dependent factor in the rate constant\cite{Borgis:1991,DMcpl:93}  
\begin{equation}
k_\text{ET} = \exp[\tfrac{1}{2}\gamma^2 \sigma_R^2] k_\text{NA}   
\label{5}
\end{equation}
with the variance  
\begin{equation}
  \sigma_R^2=\langle \delta R^2\rangle=(\beta \kappa)^{-1} .
  \label{10}
\end{equation}
The second expression in Eq.\ \eqref{10} defines the distance variance in terms of the spring constant $\kappa$ of a harmonic restraining potential; $\beta =(k_\text{B}T)^{-1}$ is the inverse temperature. The linear scaling with temperature $\sigma_R^2\propto T$, which is the static limit of the fluctuation-dissipation theorem (FDT)\cite{Crisanti:2003wf,MARCONI:2008dp} (Johnson-Nyquist noise\cite{FeynmanLecturesV1}), should make the exponential fluctuation factor in Eq.\ \eqref{5} scale linearly with $T$. However, if the squared radius of gyration $R_g^2$ is a fair gauge of $\sigma_R^2$, the temperature scaling is more complex for a folded protein.\cite{Teeter:pnas.2001} Applying the FDT temperature scaling, the reaction activation barrier gains an explicit dependence on temperature and a corresponding term enters the activation enthalpy\cite{DMcpl:93}
\begin{equation}
\Delta H^\dag = E_a + (k_\text{B}T)^2 \frac{\gamma^2}{2\kappa} ,  
\label{101}
\end{equation}
where $E_a$ is the Arrhenius activation energy. The second term in this equation was invoked to explain a large temperature-dependent kinetic isotope effect in enzymatic proton-transfer reactions.\cite{Hatcher:2007dd,Klinman:2013cw}     

The modulation of the electronic coupling through the donor-acceptor distance $R$ can have dynamical signatures as well.\cite{DMjpcb3:19}  The reason is that diffusional dynamics of the coordinate $R(t)$ with the diffusion constant $D_R$ introduces a new time scale
\begin{equation}
\tau_\gamma = (\gamma^2 D_R)^{-1} ,
\label{9}   
\end{equation}
which is the time required to diffuse through the tunneling decay distance $\gamma^{-1}$. Two relaxation processes, with the relaxation times $\tau_x$ and $\tau_\gamma$, compete in the dynamically controlled regime of electron transfer. One finds\cite{DMjpcb3:19} $k_\text{ET}\propto \tau_\gamma^{-1}$ for thin-film protein electrochemistry when the protein-electrode distance is thermally modulated through a weak binding to the electrode. In this way, non-statistical aspects of protein flexibility\cite{Klinman:2013cw} enter the pre-exponential factor of the rate constant through the time scale of the donor-acceptor diffusional dynamics.      

Here, an analytical solution is derived for the problem of donor-acceptor electron transfer when both the reaction coordinate $X(t)$ and the donor-acceptor distance $R(t)$ are dynamical variables executing diffusional dynamics in the corresponding harmonic potentials. For the coordinate $X$, this is the Marcus parabolic free energy surface.\cite{MarcusSutin} The coordinate $R$ is assumed to fluctuate in a harmonic restraining potential around the equilibrium distance $R_e$. Motions of many particles in the medium project on two collective variables $X$ and $R$ and the harmonic well $V(X,R)$ is a potential of mean force (a free energy) affected by the thermodynamic state of the medium. Likewise, many molecular conformations of the donor-acceptor complex, coupled to the medium, will project on changes of the donor-acceptor distance $R$.  

\section{Formalism}
We will introduce dimensionless coordinates $x=(X_0-X)/\sigma_X$ and $z=\delta R/\sigma_R$.  The two-dimensional harmonic well is described by the harmonic potential of $x$ and $z$
\begin{equation}
\beta V(x,z) = \tfrac{1}{2} x^2 + \tfrac{1}{2} z^2 .
\label{11}  
\end{equation}

\begin{figure}
\includegraphics*[clip=true,trim= 0cm 0cm 0cm 0cm,width=6cm]{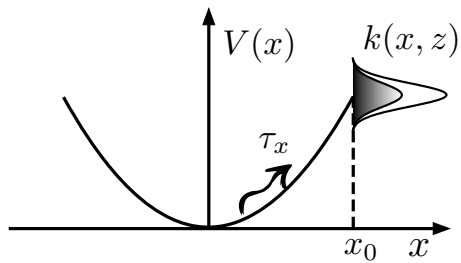}
\caption{Harmonic potential of mean force along the energy-gap coordinate $x=(X_0-X)/\sigma_X$. The sink at $x=x_0$ allows particles to escape from the well with the tunneling frequency $k(x,z)$ (Eq.\ \eqref{12}). The unshaded and shaded curves specify $k(x,z)$ at smaller and larger donor-acceptor distances, respectively. The dynamics along the $x$-coordinate are characterized by the relaxation time $\tau_x$. }
\label{fig:1}
\end{figure}

The dynamics of the probability density $n(x,z,t)$ in the reactant state is  determined by diffusion along the coordinates $x$ and $z$ and a sink of the reactant population (Fig.\ \ref{fig:1}).\cite{Sumi:86,Tang:2005io} The sink is specified by the golden rule tunneling frequency when resonance $X=0$ is reached along the energy-gap reaction coordinate
\begin{equation}
k(x,z) = \Delta_e e^{-\zeta z} \delta(x-x_0) , 
\label{12}
\end{equation}
where $\zeta=\gamma\sigma_R$ and $x_0=X_0/\sigma_X$. Further, $\Delta_e$ is given in terms of the electronic coupling $V_e$ at the equilibrium donor-acceptor separation
\begin{equation}
\Delta_e = \frac{2\pi V_e^2}{\hbar\sigma_X}  .
\label{13}
\end{equation}

The probability density $n(x,z,t)$ in two-dimensional space propagates in time by the diffusional Fokker-Planck equation.\cite{Sumi:86} This equation can be converted to the Hamiltonian form\cite{Gardiner:97}
\begin{equation}
\partial_t \bar n = - \left[H + k \right] \bar n  ,
\label{14}
\end{equation}
where $\bar n(x,z,t)$ is obtained by multiplying the probability density with the square-root of the equilibrium distribution function 
\begin{equation}
\bar n(x,z,t) = n(x,z,t)\exp\left[\tfrac{1}{2}\beta V(x,z)\right] .  
\label{15} 
\end{equation}

The Hamiltonian function in Eq.\ \eqref{14} is derived from the Fokker-Planck equation for diffusion in a two-dimensional quadratic potential\cite{Gardiner:97} and is given by the following relation 
\begin{equation}
H = H_x + H_z , 
\label{16}
\end{equation}
where
\begin{equation}
H_y = -\tau_y^{-1}\left[\frac{\partial^2}{\partial y^2} - \frac{y^2}{4} + \frac{1}{2} \right]  
\label{17}  
\end{equation}
and $y=x,z$. The relaxation time $\tau_x$ represents Stokes-shift dynamics along the reaction coordinate $X$ and $\tau_z=\tau_R=\langle \delta R^2\rangle/D_R$ is the characteristic diffusion time related to the diffusion constant $D_R$. It should be viewed as an effective diffusion constant describing dissipative translational dynamics along a single coordinate $R$. For instance, when donor and acceptor are linked by a polypeptide chain, $D_R$ is the end-to-end diffusion constant.\cite{Eaton:2021fl} When the distance dynamics can be sampled by numerical simulations, $\tau_z$ can be associated with the integral relaxation time of the normalized time correlation function $C_R(t)=\langle\delta R(t)\delta R(0)\rangle$ as
\begin{equation}
\tau_z=\int_0^\infty dt C_R(t)/C_R(0) .  
\end{equation}
  
The population dynamics specified by Eq.\ \eqref{14}  can be solved by applying the Sumi-Marcus formalism.\cite{Sumi:86} The solution is conveniently cast in terms of bra and ket vectors. The equilibrium state is specified as
\begin{equation}
\langle x,z|e\rangle = \frac{1}{\sqrt{2\pi}} \exp\left[-\tfrac{1}{2}\beta V(x,z)\right] . 
\label{18} 
\end{equation}
This definition allows the standard normalization of the inner product 
\begin{equation}
\langle e| e\rangle =\int dx dz \langle e|x,z\rangle\langle x,z| e\rangle  =  1  
\end{equation}
corresponding to a single electron in the reactant state when the equilibrium distribution is created at $t=0$.    

Assuming an equilibrium probability density of the reactants at $t=0$, the evolution of the system is described by the ket vector $|\bar n(t)\rangle$ through the following dynamic equation  
\begin{equation}
\partial_t |\bar n(t)\rangle = -(H+k) |\bar n(t)\rangle  .
\label{19}
\end{equation}
The ket $|\bar n(t)\rangle$ is defined as 
\begin{equation}
\langle x,z|\bar n(t)\rangle =  \sqrt{2\pi} \bar n(x,z,t)  . 
\label{20}
\end{equation}
This definition guarantees the initial condition $|\bar n(0)\rangle=|e\rangle$ if equilibrium is assumed at $t=0$ 
\begin{equation}
n(x,z,0)=n_\text{eq}(x,z)=(2\pi)^{-1}\exp\left[-\beta V(x,z)\right] .   
\label{21}
\end{equation}

The time Laplace transform $|\bar n(s)\rangle$ of  $|\bar n(t)\rangle$ is given by the Green's function (resolvent of the operator $H+k$)
\begin{equation}
|\bar n(s)\rangle = G(s) |e\rangle  , 
\label{22}
\end{equation}
where
\begin{equation}
G(s) = \left[s+ H + k\right]^{-1}  . 
\label{23}
\end{equation}
The Green's function of full dynamics with the sink satisfies the Dyson equation\cite{Mahan:90} 
\begin{equation}
G(s) = G_0(s) - G_0(s) k G(s)  
\label{24}
\end{equation}
in which $G_0(s)=(s+H)^{-1}$ is the Green's function of diffusional dynamics in the parabolic well $V(x,z)$. 

The time-dependent population of the reactant state $\Gamma(t)$ is given by the bra-ket
\begin{equation}
\Gamma(t)  = \langle e | \bar n(t)\rangle  
\label{25}
\end{equation}
for which the initial condition $\Gamma(0)=\langle e|e\rangle=1$ is satisfied. From Eq.\ \eqref{23}, the Laplace transform of $\Gamma(t)$ is the matrix element of the Green's function projected on the equilibrium state
\begin{equation}
\Gamma(s) =  \langle e| G(s)|e\rangle  .
\label{26}
\end{equation}

One can next define the number flux $j(t) = d\Gamma(t)/dt$ with the following result for the Laplace transform
\begin{equation}
j(s) = \langle e | sG(s) -1 |e\rangle  . 
\label{27} 
\end{equation}
The property of the equilibrium state to produce zero eigenvalue for the propagation Hamiltonian $H|e\rangle =0$, $\langle e|H= 0$ can be applied to combine Eqs.\ \eqref{24} and \eqref{27} with the result 
\begin{equation}
j(s) = -\langle e | kG(s)|e\rangle    .
\label{28} 
\end{equation}

A closed-form solution of the dynamic problem can be achieved by applying the factorization anzatz introduced by Wilemski and Fixman\cite{Wilemski:74} and, independently, by Sumi and Marcus.\cite{Sumi:86} It consists of projecting out the coupled dynamics of $G_0$ and $G$ on the equilibrium manifold\cite{Sumi:86}
\begin{equation}
G_0(s) k G(s) \to \langle k \rangle ^{-1} G_0(s) k|e\rangle \langle e| kG(s)  .
\label{29} 
\end{equation}
Here,
\begin{equation}
\langle k\rangle = \langle e| k|e\rangle   
\label{30}
\end{equation}
is the average rate constant for the population decay assuming an equilibrium distribution of the reactant configurations unperturbed by the reaction dynamics (transition-state theory). Weiss showed\cite{Weiss:84} that the Wilemski-Fixman approximation is equivalent to the factorization relation
\begin{equation}
  \langle e|(kG_0)^nk|e\rangle \rightarrow \frac{1}{\langle k \rangle^{n-1}}\left(\langle e|kG_0k)|e\rangle\right)^n,\quad n>0
\end{equation}
It is equivalent to the decoupling ansatz in Eq.\ \eqref{29}, which becomes exact\cite{Sumi:86,Weiss:84} for the diffusional dynamics with the $\delta$-function sink $k(x)\propto\delta(x-x_0)$. It is still an approximation for the problem at hand given that $k(x,z)$ is not a $\delta$-function along the $z$-coordinate. This has to be so since a $\delta$-function in higher dimensions does not affect diffusion.  

In the case of the two-dimensional potential well $V(x,z)$ one obtains for the average rate constant
\begin{equation} 
\langle k\rangle = \frac{V_e^2}{\hbar}\left(\frac{\pi}{k_\text{B}T\lambda}\right)^{1/2}\exp\left[\tfrac{1}{2}\zeta^2-\beta \Delta G^\dag\right], 
\label{31}
\end{equation}
where $\Delta G^\dag$ is the Marcus activation barrier\cite{MarcusSutin} 
\begin{equation}
\Delta G^\dag=\frac{(\lambda+\Delta G_0)^2}{4\lambda}  . 
\label{45}
\end{equation} 
When the modulation of the donor-acceptor electronic coupling is neglected ($\zeta\to 0$), Eq.\ \eqref{31} becomes the standard expression for the golden rule nonadiabatic electron transfer in Eq.\ \eqref{2}.\cite{Barbara:96} 

A closed-form solution for the Laplace-transformed number flux follows from Eqs.\ \eqref{28}  and \eqref{29}
\begin{equation}
j(s) = -\langle k\rangle \left[s+s a(s) \right]^{-1}  ,
\label{32}
\end{equation}
where
\begin{equation}
a(s) =  \langle k \rangle ^{-1} \langle e| kG_0(s)k |e\rangle .
\label{33}
\end{equation}
The problem of population dynamics is reduced to the calculation of the function $a(s)$.\cite{Weiss:84,DMjpcb3:19}  It is the Laplace transform of the time-dependent function
\begin{equation}
A(t) =   \langle k \rangle ^{-1} \langle e| k e^{-Ht} k |e\rangle 
\label{34}
\end{equation}
satisfying the condition
\begin{equation}
A(0) = \sqrt{2\pi} \langle k\rangle e^{\zeta^2+x_0^2/2}   .
\label{35}
\end{equation}

Because of the additivity in the Hamiltonian (Eq.\ \eqref{16}), the time evolution operator in Eq.\ \eqref{34} splits in the product of individual propagators over each reaction coordinate
\begin{equation}
\langle xz|e^{-Ht}|x'z'\rangle =\langle x|e^{-H_xt}|x'\rangle \langle z|e^{-H_zt}|z'\rangle .
\label{36}   
\end{equation}
Each of them is the well-established propagator of the Ornstein-Uhlenbeck stochastic process\cite{Gardiner:97} describing diffusional dynamics in a harmonic potential well
\begin{equation}
\begin{split}
  \langle x |e^{-H_xt}|x'\rangle = &\left[2\pi(1-e^{-2\tau}) \right]^{-1/2}  \\
&\exp\left[-\frac{x^2+x'^2}{4}\mathrm{coth}\tau + \frac{xx'}{2\mathrm{sinh} \tau} \right] ,
\end{split}
\label{37}
\end{equation}
where $\tau=t/\tau_x$. The same equation applies to stochastic dynamics along the $z$-coordinate upon the replacement of the relaxation time $\tau_x\to\tau_z$. 

By applying the Ornstein-Uhlenbeck propagators in Eq.\ \eqref{34}, one obtains
\begin{equation}
A(\tau) = \frac{\langle k\rangle}{\sqrt{1-\chi(\tau)^2}}\exp\left[\zeta^2 \chi(\tau)^\alpha +\frac{x_0^2\chi(\tau)}{1+\chi(\tau)}  \right] ,
\label{38}
\end{equation}
where $\chi(\tau)=\exp[-\tau]$ and $\alpha=\tau_x/\tau_z$. Since only spatial integration is involved in deriving this equation, it is directly generalized to dispersive, non-Debye dynamics by replacing $\chi(\tau)$ with a more complex function of time,\cite{Chaudhury:2006hz} which, in the case of the Stokes-shift dynamics, is produced by applying the non-Debye form of the medium dielectric constant.\cite{Hynes:86,Tang:2005io} In deriving Eq.\ \eqref{38}, the range of integration of the distance variables $z,z'$ was extended to the entire real axis $-\infty<z,z'<\infty$, which assumes that only small-amplitude oscillations occur around the equilibrium distance $R_e$. This approximation does not apply to diffusion of a polypeptide chain in the loop closure problem\cite{Toan:jp076510y} (see below). In that case, one has to limit the displacement coordinate by the contact distance $a$: $a\le R< \infty$.    

Given that $A(\infty)=\langle k\rangle$, the Laplace transform of $A(t)$ can be approximated by a sum of short-time and long-time components
\begin{equation}
a(s) = s^{-1} \langle k\rangle + g(s),  
\label{39}
\end{equation}
where $g(s)$ is the Laplace transform of $A(t)$ calculated at $t<\tau_x$
\begin{equation}
A(\tau) \simeq \frac{\langle k\rangle}{\sqrt{2\tau}} \exp\left[x_0^2/2+ \zeta^2 -\tau(\alpha\zeta^2 + x_0^2/4) \right] .
\label{40}
\end{equation}
Given that $s\simeq \langle k\rangle$ and $\langle k\rangle\tau_x \ll 1$, $g(s)$ can be taken at $s=0$ with the result 
\begin{equation}
g=g(0)= \frac{\tau_x\Delta_e}{\sqrt{x_0^2+4\alpha\zeta^2}}  e^{3\zeta^2/2} .
\label{41}
\end{equation}

In this approximation, the time-dependent flux describes the first-order population kinetics $j(t)=-k_\text{ET}e^{-k_\text{ET}t}$ with the electron-transfer rate constant given by an analog of Eq.\ \eqref{1}
\begin{equation}
k_\text{ET}=  \frac{\langle k\rangle}{1+g} , \quad \langle k\rangle = k_\text{NA} e^{\zeta^2/2}. 
\label{42}
\end{equation}
The equation for the average rate constant $\langle k\rangle$ is the static limit for the modulation of the electronic coupling by donor-acceptor vibrations as shown in Eq.\ \eqref{5}.

The rate constant $k_\text{ET}$ is identified in the present formulation with the initial decay of the reactant population initiated from the equilibrium distribution at $t=0$. The population decay becomes non-exponential if the entire function $g(s)$ is used in Eq.\ \eqref{39} and alternative definitions of the effective reaction rate constant can be applied.\cite{Sumi:86} 

Two limits can be considered in Eq.\ \eqref{41}. If $x_0^2\gg 4\gamma^2\langle(\delta R)^2\rangle (\tau_x/\tau_z)$, one obtains at $X_0\simeq \lambda$ (small driving force) 
\begin{equation}
g \simeq \frac{2\pi V_e^2}{\hbar \lambda}\tau_\text{eff},\quad \tau_\text{eff}=\tau_x e^{3\gamma^2\langle (\delta R)^2\rangle/2}    .
\label{43}
\end{equation}
If fluctuations of the donor-acceptor distance are neglected ($\zeta\to 0$), this result is consistent with previous calculations of the solvent dynamic effect on electron transfer\cite{Sumi:86,Rips:1987qf} (note, however, that the factor $4\pi\tau_L$ appears in Ref.\ \onlinecite{Rips:1987qf} instead of $2\pi \tau_\text{eff}$ here; $\tau_L$ is the longitudinal dielectric relaxation time). In the solvent-controlled limit, Eqs.\ \eqref{42} and \eqref{43} lead at $\zeta\to 0$ to the result identical to that listed by Sumi and Marcus (Eq.\ (8.8) in Ref.\ \onlinecite{Sumi:86}) and Hynes (Eq.\ (3.12) in Ref.\ \onlinecite{Hynes:86}) 
\begin{equation}
k_\text{ET} =\tau_x^{-1} \sqrt{\pi^{-1}\beta \Delta G^\dag}   e^{-\beta \Delta G^\dag},
\label{44}
\end{equation}
where $x_0^2/2=\beta \Delta G^\dag$ and the reaction activation barrier is given by Eq.\ \eqref{45}. 

In the opposite limit $x_0^2\ll 4\gamma^2\langle(\delta R)^2\rangle (\tau_x/\tau_z)$, one gets 
\begin{equation}
 g \simeq \frac{\pi V_e^2}{\hbar \gamma\sigma_X\sigma_R}\tau_\text{eff},\quad \tau_\text{eff}=\sqrt{\tau_x\tau_z}  e^{3\gamma^2\langle (\delta R)^2\rangle/2}    .
 \label{46}
\end{equation}
This limit, corresponding to sufficiently small $x_0$, also eliminates the unphysical divergence of $g$ at $x_0\to 0$ appearing in solutions based on the first passage time.\cite{Friedman:1982eu} This divergence is not real since Eq.\ \eqref{41} has been derived under the assumption of $\langle k\rangle \tau_x\ll 1$, which requires a sufficiently large activation barrier $\beta\Delta G^\dag=x_0^2/2$ (Eq.\ \eqref{45}).  However, avoiding this singularity by analytical means provides a more broadly applicable and mathematically stable solution for the crossover parameter. The general solution for the crossover parameter $g$ is somewhat cumbersome, but includes both limits of Eqs.\ \eqref{43} and \eqref{46}  
\begin{equation}
g =   \frac{2\pi \tau_x V_e^2 }{\sigma_X\hbar}\frac{e^{3\gamma^2\langle (\delta R)^2\rangle/2} }{\sqrt{2\beta \Delta G^\dag + 4\tau_x/\tau_\gamma}}.
\label{47}
\end{equation}
where $\tau_\gamma$ is from Eq.\ \eqref{9} and $2\beta \Delta G^\dag=(\lambda+\Delta G_0)^2/\sigma_X^2$ (Eq.\ \eqref{45}). 

We find that the effective relaxation time $\tau_\text{eff}$ in the solvent-controlled regime of electron transfer is specified by the geometric mean of relaxation times $\tau_x$ and $\tau_z$. This result is distinct from the solution of a similar problem for electrode kinetics\cite{DMjpcb3:19} where $k_\text{ET}\propto \tau_\gamma^{-1}$ (Eq.\  \eqref{9}) was derived for $x_0^2\ll 4\tau_x/\tau_\gamma$.

\section{Discussion} 
Theories of solvent dynamical effect on electron transfer allow a turnover in the pre-exponential factor of the rate constant between the golden rule tunneling expression, with the rate constant proportional to $V_e^2$, and Kramers-type kinetics,\cite{Kramers:1940wm} with the rate constant proportional to the reciprocal relaxation time of the medium. The dynamical crossover parameter $g$ in Eqs.\ \eqref{1} and \eqref{42} is proportional to the product of $V_e^2$ and the medium relaxation time (Eq.\ \eqref{6}). The relaxation time is experimentally and computationally defined by the Stokes-shift time correlation function and can be approximated by the longitudinal dielectric relaxation time of the solvent\cite{Bagchi2012} $\tau_L$ when Stokes-shift dynamics are calculated from dielectric theories.\cite{Rips:1987qf,Bagchi2012} These dynamics are often complex, involving both the short-time ballistic and long-time collective components.\cite{Jimenez:94,Gayathri:96} 

The solution of the Sumi-Marcus two-dimensional diffusion problem\cite{Sumi:86} presented here incorporates diffusional dynamics of the donor-acceptor distance into the rate constant of electron transfer. While the solution is general, applications to protein electron transfer\cite{Winkler:2014ds} seem to be particularly relevant. Two-dimensional diffusion occurs in the potential of mean force $V(x,z)$, which implies that it is constructed by projecting all medium modes that are faster than the reaction kinetics on two collective coordinates. Time separation is required for media with dispersive dynamics, such as proteins. A substantial manifold of conformational modes of a protein is slower than the time of a typical electron-transfer reaction and $V(x,z)$ is by necessity a partial (nonergodic) free energy surface. What it practically means is that the parabolic force constants, specified by $\lambda$ and $\langle(\delta R^2 \rangle$, are nonergodic parameters referring to the range of medium frequencies exceeding $k_\text{ET}$.\cite{DMjpcm:15,DMsm:17} 

\begin{figure}
\includegraphics*[clip=true,trim= 0cm 2cm 0cm 0cm,width=9cm]{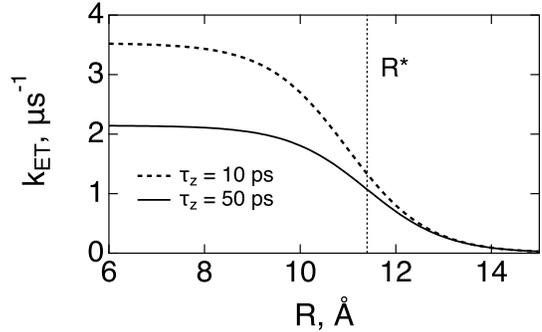}
\caption{Distance dependence of the rate constant of electron transfer with the parameters typical for protein electron transfer:\cite{Winkler:2014ds} $\lambda=0.8$ eV, $\gamma=1.2$ \AA$^{-1}$, $\langle(\delta R)^2 \rangle=1$ \AA$^2$. The electronic coupling in Eq.\ \eqref{4} is adopted\cite{Hopfield:1974uq} with $V_e=5$ meV and $R_e=7$ \AA. The two lines are drawn with $\tau_x=100$ ps and $\tau_z=10$ ps (dashed) and $50$ ps (solid). The vertical dotted line indicates the crossover distance $g(R^*)=1$ for $\tau_z=50$ ps. }
\label{fig:2}
\end{figure}

The crossover from solvent control to tunneling control leads to a step-wise shape of the rate constant as a function of the donor-acceptor distance: the rate constant is independent of the distance at short separations, followed by an exponential decay due to the falloff of the tunneling probability at longer distances (Fig.\ \ref{fig:2}). This phenomenology has been confirmed for electrochemical reactions of surface-immobilized proteins.\cite{Wei:2004cp}  

The turnover from tunneling to solvent control implies a change in the Arrhenius slope of the rate constant. As the relaxation time becomes longer with lowering temperature, the activation enthalpy switches from the nonadiabatic limit in Eq.\ \eqref{101} to a new value in the solvent-controlled regime
\begin{equation}
\Delta H^\dag = E_a+E_\tau - (k_\text{B}T)^2 \frac{\gamma^2}{\kappa}  ,
\label{102}
\end{equation}
where $E_\tau$ is the activation energy of the relaxation time. Eqution \eqref{101} thus refers to the Arrhenius slope at high temperatures (h) while Eq.\  \eqref{102} gives the slope at lower temperatures (l). The change in the slope $\Delta\Delta H^\dag = \Delta H^\dag(l) - \Delta H^\dag(h)=E_\tau - 3(k_\text{B}T\gamma)^2/(2\kappa)$ depends on relative magnitudes of two terms in this equation.     

The crossover distance $R^*$, $g(R^*)=1$ turned out to be exceptionally large, $R^*\simeq 14$ \AA, for thin-film protein electrochemistry.\cite{Wei:2004cp,Winkler:2014ds} It nearly coincides with the largest separation, $\simeq 14-15$ \AA, within which most activated electron-transfer reactions are found in biological energy chains.\cite{Page:03,Gray:ChemSci2021} A large turnover distance comes as a result of ``modulating'' donor-acceptor vibrations to contrast with  ``promoting'' vibrations\cite{Antoniou:2006,Hay:nchem.2012} proposed to drive enzymatic reactions. The dominance of slow distance dynamics and a large value of the distance variance in a soft harmonic potential restraining the protein at the electrode\cite{DMjpcb3:19,Zitare:2020fl} yield large values of $R^*$. 

One of the principal results of the proposed solution is a significant enhancement of the effective relaxation time entering the crossover parameter $g$ (Eqs.\ \eqref{6} and \eqref{47}). The enhancement factor, $\exp[3\gamma^2\langle (\delta R)^2\rangle/2]$ in Eqs.\ \eqref{43} and \eqref{46}, carries an exponential dependence on the variance of the donor-acceptor separation $\langle (\delta R)^2\rangle$, which allows protein elasticity to affect electron transfer.  The effective relaxation time exceeds $\tau_L$ by many orders of magnitude. Flexible donor-acceptor motifs, such as protein loops and peptide chains, and soft binding of electron-transfer proteins and cofactors to larger protein complexes will lead to redox reactions that show a greater tendency to fall in the solvent-controlled regime. These reactions must demonstrate a wider range of donor-acceptor separations with distant-invariant rate constants. Electronic coupling is irrelevant for these reactions, and the rate constant's pre-exponential factor is controlled by the protein dissipative dynamics.  

Saturation of the rate constant at shorter donor-acceptor separations suggests an important design principle: cofactors in electron-transfer chains are best placed near the crossover distance $R^*$ to maximize both the rate of electron transport and the distance travelled. Alternatively, a rigid environment is essential\cite{SacquinMora:2007jm} for a given donor-acceptor separation to avoid reduction of the rate constant by the crossover parameter $g$ in Eq.\ \eqref{1}. This notion applies only to the active site since the ability to maintain flexibility of the protein-water interface is critical for function.\cite{Weltz:jacs2020} 

Another parameter entering the exponential factor in the effective relaxation time is the tunneling decay length $\gamma^{-1}$. The parameter $\gamma$ is much larger in magnitude for proton transfer ($\simeq 60$ \AA$^{-1}$ in Ref.\ \onlinecite{Borgis:1991} and $\simeq 46$ \AA$^{-1}$ in Ref.\ \onlinecite{Hatcher:2007dd}) compared to electron transfer ($\simeq 1.2$ \AA$^{-1}$) and the solvent dynamical control discussed here might be relevant for these reactions. The variance is relatively small, $\langle (\delta R)^2\rangle\simeq 10^{-2}$ \AA$^2$, for proton transfer, but $3\gamma^2\langle (\delta R)^2\rangle/2\simeq 38$, entering the exponential function, is sufficiently large to modify the standard predictions of the noadiabatic tunneling theory. Kiefer and Hynes\cite{Kiefer:2004dk} and Cui and Karplus\cite{Cui:jacs2002} indicated that the barrier for proton transfer is substantially affected by donor-acceptor vibrations, but their analysis does not include the solvent-controlled regime. 

The Stokes-shift dynamics and the dynamics of the donor-acceptor distance are described here in the Markovian  approximation excluding memory effects from the corresponding Langevin equation. Extensions to non-Debye polarization dynamics\cite{Tang:2005io} are possible as suggested by Hynes.\cite{Hynes:86} The use of Markovian dynamics for the donor-acceptor distance is justified for reaction times on the scale of 1--100 nanoseconds when the distance dynamics are still biphasic, with two exponential decays carrying relaxation times in the picosecond and nanosecond domains.\cite{DMpb:12} For the reaction times in the range of milliseconds and longer, more complex distance dynamics reported by single-molecule measurements\cite{Min:05} become relevant. Power-law tails of the memory function $\propto t^{-1/2}$ were identified in the time window $10^{-3}-10$ s.\cite{Min:05} Such stretched memory kernels project on a time-dependent diffusion coefficient $D_R(t)$ in the Fokker-Planck equation.\cite{Chaudhury:2006hz}  The present formalism still applies if the scaled time\cite{Hynes:86,Tang:2005io,Satija:2017ew} $t\propto \int_0^t D_R(t')dt'$ is used in the Ornstein-Uhlenbeck propagator (Eq.\ \eqref{37}) for the $R$-coordinate. Such a theory extension is likely not needed for electron-transfer reactions in the nanoseconds to microseconds reaction time window typical for biological energy chains.\cite{Bioenergetics3} Note, however, that heterogeneity of slowly-exchanging conformations,\cite{English:nchembio759} each allowing a separate electron-transfer channel, will require a corresponding average of the population flux
\begin{equation}
\langle j(t) \rangle_\text{het} = \partial_t \left\langle e^{-k_\text{ET} t}\right\rangle_\text{het}  , 
\end{equation}
where $\langle \dots \rangle_\text{het}$  denotes an average over the slow conformational variables. 

There is a clear connection between the problem considered here and a much studied problem of loop closure for polypeptide chains.\cite{Szabo:1980aa,Lapidus:2002gd,Toan:jp076510y,Eaton:2021fl}  The dynamics are monitored by electron-transfer quenching between the photoexcited donor and acceptor at the ends of the chain.\cite{Lapidus:2002gd,Eaton:2021fl} The canonical solution of the problem by Szabo, Schulten, and Schulten (SSS)\cite{Szabo:1980aa} separates the observed kinetics into the electron-transfer quenching rate $k_q$ and the first-passage time $\tau_\text{FP}$ of diffusional arrival to the quenching distance. The inverse of the total observable rate constant $k_\text{obs}$ is a sum of the time of quenching and time of diffusional first passage 
\begin{equation}
\frac{1}{k_\text{obs}} = \frac{1}{k_q} + \tau_\text{FP} .  
\label{50}
\end{equation}

The difference between the SSS formulation and the present model is that here, instead of assuming a given quenching rate, both distance diffusion and medium fluctuations driving the system across the activation barrier are parts of the same fluctuating medium projecting its thermal motions on two collective coordinates. The time scales of diffusional and polarization fluctuations are viewed as similar, which applies to compact structures of folded proteins. To connect to the SSS model, one can write Eq.\ \eqref{42} in the form
\begin{equation}
\frac{1}{k_\text{ET}} = \frac{1}{\langle k\rangle} + \sqrt{\frac{\pi}{2}\tau_\gamma\tau_x} e^{\gamma^2 \langle(\delta R)^2\rangle +\beta \Delta G^\dag }   .
\label{51}
\end{equation}

It is obvious that $k_q=\langle k\rangle$, but the structure of the second term is different from the SSS model. As mentioned above, the solution leading to Eq.\ \eqref{51} was obtained by assuming small oscillations of the donor-acceptor complex near the equilibrium distance $R_e$ and thus not considering the diffusional propagation to the quenching contact configuration. What is clear, however, is that diffusional and reaction dynamics do not decouple when the tunneling probability carries an exponential distance dependence. This seems to be a general result. Indeed, Makarov and co-workers have previously noticed this deficiency of the SSS solution.\cite{Cheng:jp902291n}  The decoupling between reaction and diffusion dynamics assumed in the SSS model does not apply for a distant-dependent quenching rate: there is always a range of distances with similar diffusion and quenching times. 

The data for the loop closure kinetics are often analyzed\cite{Lapidus:physrevlett.87.258101,Lapidus:2002gd} in terms of the dependence of $k_\text{obs}^{-1}$ on the medium viscosity $\eta$ under the expectation that $\tau_\text{FP}\propto\eta$ and the quenching kinetics are not affected by the medium viscosity. Similarly to these expectations, $g\propto\eta$ is predicted in both Eqs.\ \eqref{43} and \eqref{46}, while a more complex functionality, often fitted by a power law,\cite{DMjpcb3:19,Zitare:2020fl} might be realized in the intermediate regime described by Eq.\ \eqref{47}. Two time scales, $\tau_x$ and $\tau_z$, compete in the effective relaxation time in the intermediate regime and the power law is used as an empirical tool in the absence of an established functionality. Omitting these complications, the second term in Eq.\ \eqref{51} mostly scales linearly with $\eta$. 

The rate constant $k_\text{ET}$ in the present formulation refers to a stable, bound donor-acceptor complex. If the donor-acceptor complex is formed by diffusional encounter, a separate step of diffusional kinetics (second term in Eq.\ \eqref{50}) needs to be added to the analysis of kinetic data. Alternatively,  the two-dimensional diffusional equations need to be extended to realistically model propagation to the contact distance. Adopting the phenomenological first route, the observable rate constant becomes
\begin{equation}
\frac{1}{k_\text{obs}} = \frac{1}{\langle k\rangle} + \sqrt{\frac{\pi}{2}\tau_\gamma\tau_x} e^{\gamma^2 \langle(\delta R)^2\rangle +\beta \Delta G^\dag }  + \tau_\text{FP} .
\label{52}
\end{equation}
The inverse rate constant $k_\text{ET}^{-1}$ from Eq.\ \eqref{51} thus replaces $k_q^{-1}$ in Eq.\ \eqref{50}. The last two terms, linear in $\eta$, add up in Eq.\ \eqref{52} when the dependence of $k_\text{obs}$ on the solvent viscosity is measured. 

To conclude, the main result of the present theory is that the electron-transfer rate is affected by the protein dynamics and flexibility through the effective relaxation time $\tau_\text{eff}$ in Eq.\ \eqref{46}. It involves the dynamical component through the geometric mean of the Stokes-shift and oscillation relaxation times and the effect of protein elasticity through the displacement variance in the exponential factor.

\acknowledgements 
This research was supported by the Army Research Office (ARO-W911NF2010320) and by the National Science Foundation (CHE-2154465). 

\section*{Author Declarations}
\textbf{Conflict of Interest}\\
The author has no conflicts to disclose.

\section*{DATA AVAILABILITY}

The data that support the findings are available from the author upon request.

\bibliography{chem_abbr,dielectric,dm,statmech,protein,et,bioet,liquids,solvation,dynamics,simulations,surface,water,glasset,nano,viscoelastisity,conductivity,ions,diffusion,glass,photosynthNew,biophys,pt,bioenergy,enm,pt}

\end{document}